\documentstyle[prl,psfig,aps,multicol]{revtex}
\psfigurepath{/month/saif/Part2/latex/}

\begin{document}
\draft
\title{Dynamical Revivals in Fermi Accelerator Model}
\author{Farhan Saif}
\address{Department of Electronics, Quaid-i-Azam University, Islamabad, Pakistan.}
\date{\today}
\maketitle

\begin{abstract}
We investigate quantum revivals in the dynamics of an atom in an atomic 
Fermi accelerator. It is
demonstrated that the external driving field influences the revival time
significantly. Analytical expressions are presented which are based on
semiclassical secular theory. These analytical results explain the
dependence of the revival time on the characteristic parameters of the
problem quantitatively in a simple way. They are in excellent agreement with
numerical results.
\end{abstract}


\begin{multicols}{2}
\narrowtext

Periodically driven quantum systems
have received considerable attention over the past
few years due to the presence of Anderson-like 
localization~\cite{kn:benv1}.
Some recent numerical work indicates the presence of 
quantum revivals~\cite{kn:hogg} in periodically driven quantum
systems~\cite{kn:haak,kn:bres} as well as in other two-degree-of-freedom
systems~\cite{kn:toms}. Though many aspects of quantum localization are well understood by now,
so far even the most elementary questions concerning 
revival effects in periodically driven quantum systems 
are not yet comprehensible. Under which circumstances do such revival
phenomena appear and how do these
revival effects depend on the frequency and on the driving amplitude of
the applied periodic force? 
The main intention of this letter is to address these questions 
by considering a physical system which is of 
particular interest in the field of 
atom optics~\cite{kn:mlyn,kn:pill,kn:arim}, namely an atom in 
an atomic Fermi accelerator. It is made up of an
atomic mirror~\cite{kn:wallis} modulated by means of a external periodic
field~\cite{kn:sten,kn:saif1}. 
It will be shown that quantum revivals can be 
observed in this physical system and that the dominant influence
of the periodic driving force results in a change of the revival time.
Simple analytical results are presented which explain the
quantitative dependence of the revival time on the driving frequency 
and the strength of the driving amplitude. These analytical
results are in excellent agreement with numerical results.

Let us consider an atom in a periodically
driven gravitational cavity~\cite{kn:wallis}.
The atom moves under the influence of gravity 
in the positive $\tilde z$-direction 
and is reflected back as it hits a mirror, as shown in Fig~\ref{fg:model}. 
The atomic mirror is assumed to be 
made up of an evanescent electromagnetic field with electric field
strength ${\it{\bf E}}(\tilde z, t) = {\bf e}_x {\it E}_0 e^{-\omega_L \tilde z/c}
e^{-i\omega_L  t} + {\rm c.c}$.
In order to study the effect of an external driving force
on this elementary quantum system the evanescent wave is assumed to
be modulated by an acusto-optical modulator~\cite{kn:sten},
that is  ${\it E}_0={\cal E}_0\exp(\epsilon \sin\omega t)$.
Assuming that the optical driving frequency $\omega$ is well detuned
from any atomic resonance and taking into account the symmetry of the
problem in the $(x,y)$-plane the effective one-dimensional
center-of-mass motion of the atom is governed 
by the Hamiltonian~\cite{kn:saif1}
\begin{equation}
H=\frac{{p}^2}{2M} + Mg{\tilde z}
+ \frac{\hbar \Omega_{eff}}{4} e^{-2 \omega_L{\tilde z}/c+\epsilon\sin\omega t}.
\label{ham}
\end{equation}
Here ${p}$ is the atomic center-of-mass momentum, 
$M$ is the atomic mass and $g$ denotes the gravitational
acceleration. The effective Rabi frequency $\Omega_{eff}$ 
characterizes the strength of the influence of the applied electric 
field.

The potential generated by the gravitational acceleration
and by the evanescent laser field has the
approximate form of a time dependent triangular potential
well like in the Fermi accelerator~\cite{kn:fermi}.
Thus, in the subsequent, approximate
treatment we replace this potential  by the idealized, simpler form
of a triangular well with an infinitely high potential barrier at
the position $\tilde z=\frac{c\epsilon}{2\omega_L} \sin\omega t$ 
of the mirror.
In this approximation our results become independent of details
of the evanescent laser field, such as the laser frequency $\omega_L$
and the value of the effective Rabi frequency $\Omega_{eff}$.

In the moving coordinate we may write the time dependent
Schr\"odinger equation as~\cite{kn:saif2}
\begin{equation}
i \hbar\dot\chi 
=\left\{\frac{p^2}{2M} + Mz(g-\lambda\omega^2\sin\omega t) 
+ V_0 e^{-\kappa z}\right\}\chi\;,
\label{schr1}\end{equation}
where $z=\tilde z- \lambda\sin\omega t$. Here we are 
considering $V_0\equiv\hbar\Omega_{eff}/4$, 
the steepness $\kappa\equiv 2\omega_L/c$ and  the
modulation strength $\lambda\equiv c\epsilon/(2\omega_L)$.

A convenient way of obtaining insight into the 
influence of the acusto-optical external
driving force on the atomic dynamics is to
investigate the time evolution of
an atomic center-of-mass wave packet.
For this purpose we consider a Gaussian wave packet 
$\psi(0)=(2\pi\Delta z^2)^{-1/4}
\exp\{-(z-z_0)^2/(2\Delta z)^2\}\exp(-ip_0(z-z_0)/\hbar)$
at $t=0$ and propagate it in the gravitational cavity
for different modulation strength $\lambda$.
Here $z_0$ describes the average position, $p_0$ denotes
the average momentum of the wave packet and $\Delta z$ is the
spatial uncertainty. 
In Fig.~\ref{fg:reviv}  characteristic time dependences of the
autocorrelation function ${\cal C}(t)\equiv\langle\psi(0)|\psi(t)\rangle$
of the wave packet are shown for $z_0=20.1\mu m$, $\Delta z=0.28\mu m$ and $p_0=0$. 
Without external periodic perturbation (uppermost figure)
the well known scenario of revivals and fractional revivals~\cite{kn:aver} 
is clearly apparent. In the presence of 
a sufficiently weak external periodic
driving force the revivals and fractional revivals are still observable.
However, the revivals decrease in magnitude and the revival
time exhibits a pronounced
dependence on the external driving force. 

We~\cite{kn:saif2} calculate the revival time with the help
of semiclassical secular theory. 
In the vicinity of the $N$-th primary resonance the
classical  Hamiltonian corresponding to Eq.~(\ref{schr1})
can be expressed in action and angle variables ($I,\varphi$)~\cite{kn:lieb} as 
\begin{eqnarray}
H=\frac{H''}{2}(I-I_0)^2+H'(I-I_0)\nonumber\\
+ H_0(I_0) +\lambda V\sin(N\varphi-\omega t)\;.
\label{eq:reshsc}
\end{eqnarray}
Here, $I_0$ is the classical action associated with the initial
condition and
$V=(Mg)(I_0/I_N)^{2/3}$. The classical action at the 
center of the $N$-th primary resonance is
denoted by $I_N$ . 
Moreover $H'$ and $H''$ are the first and second derivatives of 
energy with respect to action calculated at $I_0$. Similarly
$H_0(I_0)$ is the energy of the unperturbed system for the initial
action $I_0$.
In Eq.~(\ref{eq:reshsc}) we have averaged 
out the fast oscillating terms so that $H(I-I_0)$ represents
an integrable one-degree-of-freedom physical system.

Introducing the transformation $N\varphi-\omega t=2\theta+\pi/2$
and quantizing the dynamics around the $N$-th resonance 
by using $I-I_0=\frac{\hbar}{i}\frac{\partial}{\partial\varphi}
=\frac{N\hbar}{2i}\frac{\partial}{\partial\theta}$~\cite{kn:berry},
the time independent Schr\"odinger equation becomes
\begin{eqnarray}
\left[-\frac{N^2H''{\hbar}^2}{8}\frac{\partial^2}
{\partial\theta^2}+\frac{\hbar}{2i}(NH'-1)
\frac{\partial}{\partial\theta} + H_0(I_0)\right.\nonumber\\
\left. +\lambda V\cos 2\theta\right]\psi={\cal E}_n\psi\;.
\label{eq:schham}
\end{eqnarray}
Thus the quasi energies ${\cal E}_n$ are determined 
by Eq.~(\ref{eq:schham}).
Here $\psi(\theta)$ has to fulfill the periodic boundary condition.
It is straightforward to write Eq.~(\ref{eq:schham}) in the form of 
a Mathieu equation by substituting 
$\psi=\phi\exp\left(-2i(NH'-1)\theta/(N^2H''\hbar)\right)$,
namely
\begin{equation}
\left[\frac{\partial^2}{\partial\theta^2} +a -2q\cos 2\theta\right]\phi=0\;,
\label{eq:math}
\end{equation}
with
\begin{eqnarray}
a&=&\frac{8}{N^2H''(I_0){\hbar}^2}\left[{\cal E}_n-H_0(I_0)
+ \frac{(NH'(I_0)-1)^2}{2N^2H''(I_0)}
\right]\;,\\
q&=&\frac{4\lambda V}{N^2H''(I_0){\hbar}^2}\;.
\end{eqnarray}
The quasi-energy eigenvalues ${\cal E}_n$ are determined 
by the solutions of Eq.~(\ref{eq:math}) and the 
requirement that $\phi(\theta+\pi)=\phi(\theta)$.
The $\pi$-periodic solutions of Eq.~(\ref{eq:math}) correspond to 
even functions of the Mathieu equation whose corresponding
eigenvalues are real~\cite{kn:abra}. 
These solutions are $\phi_{\nu}(\theta)=e^{i\nu\theta}P_{\nu}(\theta)$
where $P_{\nu}(\theta)$ is the even order Mathieu function.
In order to obtain a $2\pi$-periodic solution in $\varphi$-coordinate
we require the coefficient of $\varphi$ to be equivalent to  
an integer number. This requirement
provides us the oppertunity to find the value for 
the index $\nu$ of the Mathieu functions as 
\begin{eqnarray}
\nu=\frac{2}{N\hbar}\left[I-4I_0
+3I_0\left(\frac{I_0}{I_N}\right)^{1/3}\right]\;,
\label{eq:denu} 
\end{eqnarray}
where $I=(n+3/4)\hbar$. 
The quasi energy of the system is finally given by
\begin{equation}
{\cal E}_{n}=\frac{N^2H''{\hbar}^2}{8}a_{\nu(n)}(q)
-\frac{(NH'-1)^2}{2N^2H''}
+H_0(I_0)\;,
\label{eq:qen}
\end{equation}
with the Mathieu characteristic parameter
$a_{\nu(n)}(q)$.

In order to check this result we study case of zero modulation
strength, that is $\lambda=0$ which implies $q=0$. In this case 
the value for Mathieu characteristic parameter becomes
$a_{\nu}(q=0)=\nu^2$. This leads us to
Hamiltonian corresponding
to schr\"odinger equation of Eq.\ref{eq:reshsc} in the 
absence of external modulation. 

We~\cite{kn:saif2} employ the quasi-energy spectrum ${\cal E}_n$ 
of the driven system to evaluate the revival time. This concept 
helps us to define the time of revival $T_{\lambda}$ 
for time dependent system as
\begin{eqnarray}
T_{\lambda} &=& {4\pi \hbar}\left( {\mid
\frac{\partial^2{\cal E}_n}{\partial n^2}
\mid_{n=n_0}}\right)^{-1}.
\label{T}
\end{eqnarray}

From the quasi-energies of Eq.~(\ref{eq:qen})
we can determine the revival time in the presence of the external time 
dependent field. Considering $q<1$ 
we expand the Mathieu characteristic parameter $a_{\nu(n)}(q)$
up to $q^2$. We simplify our result by noting that
$H''=-(Mg^2)^{1/3}(\pi/9I_0^2)^{2/3}$ and $N=(\omega^3/Mg^2)^{1/3}(3I_N/\pi^2)^{1/3}$. 
Thus we finally obtain
\begin{equation}
T_{\lambda}=T_0\left[ 1-\frac{1}{2}\left\{\frac{8\lambda}
{{\hbar}^2} E_N \left( \frac{I_0}{I_N}\right)^2\right\}^{2}
\frac{3\nu^2+1}{(\nu^2-1)^3}\right].
\label{eq:rtspt}
\end{equation}
On substituting the value of $\nu$, calculated at $I=I_0$, in
Eq.~(\ref{eq:rtspt}) we get
\begin{equation}
T_{\lambda}=
T_0\left[ 1-\frac{1}{8}\left\{
    \frac{M\lambda g}{E_{n_0}} \right\}^{2}
    \frac{3(1-r)^2+a^2}{((1-r)^2-a^2)^3}\right]
\label{eq:rtsptn1} 
\end{equation}
where $r\equiv(E_N/E_{n_0})^{1/2}$ and $a\equiv r^2 \hbar\omega/4E_{n_0}$.
If the initial energy is large, {\it i.e.} $E_{n_0}\gg \hbar\omega$, we 
may consider $a^2$ much smaller than $(1-r)^2$, which leads to
\begin{equation}
T_{\lambda}=T_0\left[ 1-\frac{3}{8}\left\{ 
\frac{M\lambda g}{E_{n_0}} \right\}^{2} 
\frac{1}{(1-r)^4}\right].
\label{eq:rtsptn} 
\end{equation}

Our analytical result explain in a simple
way the quantitative dependence of the revival time $T_{\lambda}$
on the characteristic parameters of the problem, namely
the driving frequency $\omega$ and the driving amplitude $\lambda$.
In order to access the accuracy of these perturbative results
we calculate the revival time $T_{\lambda}$ by integrating the Schr\"odinger 
equation Eq.~(\ref{schr1}) numerically, and compare it with the analytically
obtained result of Eq.~(\ref{eq:rtsptn}). 
For this comparison we have considered
two different initial conditions of the atomic
wave packet above 
the surface of the atomic mirror. In Fig.~\ref{fg:revival} 
solid lines with circles corresponds to $(a)$
$z_0=29.8\mu m$ which implies a state with mean principle 
quantum number $n_0=322.51$, however, 
solid line with squares corresponds to $(b)$
$z_0=20.1\mu m$ which implies $n_0=176.16$.
In both cases initial average momentum $p_0=0$.
The first initial condition lies further away from
the center of the corresponding primary resonance as compared to the
second one.  

Numerically we find that the change in revival time depends
quadratically on the strength of external modulation $\lambda$
as predicted from Eq.~(\ref{eq:rtsptn}).
We plot the analytically obtained result for the two initial condition
in Fig.~\ref{fg:revival} with the help of dashed lines. We find that
the change of the revival time is smaller in the case $(a)$
 than in the case $(b)$. 
Our analytical result explain this dependance:
In case (a) of Fig.~\ref{fg:revival} our 
chosen initial condition $z_0=29.8\mu m$
has a higher energy $E_{n_0}$ than in case (b) 
where $z_0=20.1\mu m$.
Since the change in revival time has inverse dependence
on the square of the energy $E_{n_0}$, as a result, we observe
a smaller change in revival time when the average energy is larger as in
case $(a)$ and a larger change in revival time when the average energy 
is smaller as in $(b)$.

We have demonstrated that the dynamics of a material wave packet
in a periodically driven gravitational cavity exhibits quantum 
mechanical revivals. We have presented the first
results on the quantitative dependence of the revival time
on the characteristic parameters of the problem, namely 
the driving frequency and the driving strength.
We show that these dependences
can be understood quantitatively in a satisfactory way by
using semiclassical secular theory. 
In view of the recent experimental~\cite{kn:sten} developments 
the presented quantitative predictions 
are accessible to experimental observation.

We thank G. Alber, I.\ Bialynicki-Birula, M.\ Fortunato, 
M.\ El Ghafar, R.\ Grimm, V. Savichev, P.\ T\"orm\"a,
W. P. Schleich and A.\ Zeiler
for many fruitful discussions. The author convays his special 
thanks to G. Alber for careful look on manuscript.

\end{multicols}
\begin{figure}[tbp]
\caption{Experimental setup of atomic Fermi accelerator: Cloud of cesium
atoms is trapped and cooled in a magneto optical trap (MOT) down to the
micro-Kelvin scale. The MOT is placed at a certain height above the
evanescent wave mirror. The mirror for the atoms results due to the total
internal reflection of the laser light field from the surface of the glass
prism. The atomic mirror is modulated by providing an intensity modulation
to the evanescent light field via an acousto optic modulator. The
polarization of the evanescent wave field is inside the plane of the
reflection. We consider the atomic dynamics only along the $z$-axis, that
is, along the normal to the surface of the mirror.}
\label{fg:model}
\end{figure}

\begin{figure}[tbp]
\caption{Revival phenomena in a gravitational cavity in the absence and in
the presence of periodic modulation: Autocorrelation of a Gaussian wave
packet of a cesium atom as a function of time, prepared at $t=0$ with $%
\Delta z=0.28\mu m$, for $\lambda=0$ (a), $\lambda=0.56 \mu m$ (b), $%
\lambda=1.13 \mu m$ (c), and $\lambda=2.26\mu m$ (d). The parameters are $%
\omega=2\pi\times 0.93$KHz, $\Omega_{eff}=23.38$KHz, and $\kappa^{-1}=0.57\mu m$.
The average position of the wave packet in the gravitational cavity is $%
z_0=20.1\mu m$ which corresponds to the mean quantum number $n_0=176.16$. }
\label{fg:reviv}
\end{figure}

%\begin{figure}[tbp]
%\caption{Revival times $T_{\lambda }$ as obtained from exact numerical
%calculations of the time dependent Schr\"{o}dinger equation Eq.~(\ref{schr1}%
%) (sold lines), and from analytical results based on Eqs.~(\ref{revpert})
%(dotted lines) and ~(\ref{eq:rtspt}) (dashed line): Initial conditions are
%(a) $z_{0}=29.8\mu m$, $r=0.87$ and (b) $z_{0}=20.1\mu m$, $r=0.77$. The
%other parameters are the same as in Fig.~\ref{fg:reviv}. Here we have $\bar{%
%\lambda}=\lambda \omega ^{2}/g$. }
%\label{fg:plot}
%\end{figure}

\begin{figure}[tbp]
\caption{We compare the ratio between revival time for driven system and
revival time of undriven system, $T_{\lambda}/T_0$, for two different
initial conditions, $z_0=29.8\mu m$ (solid line with circles) and $%
z_0=20.1\mu m$ (solid line with squares). The comparison comes from
exact numerical calculations of time dependent Schr\"{o}dinger equation
Eq.~(\ref{schr1}) (solid lines), and from
analytical results based on
Eq.~(\ref{eq:rtsptn1}) (dashed lines). 
All the other parameters are the same as
in Fig.~\ref{fg:reviv}. }
\label{fg:revival}
\end{figure}

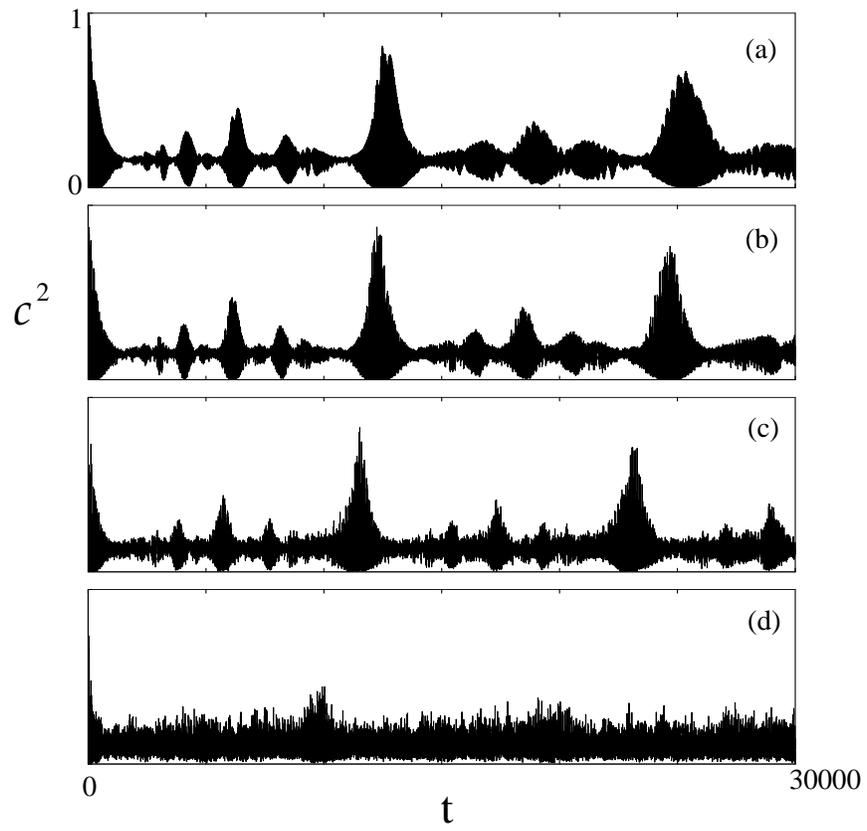
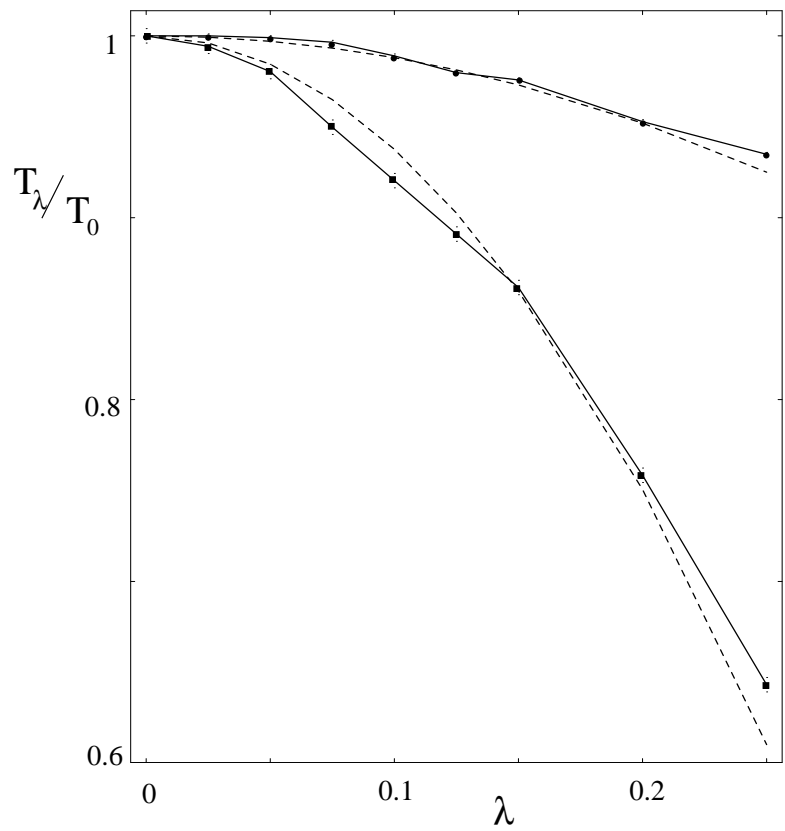
\end{document}